# Imaging objects through scattering layers and around corners by retrieval of the scattered point spread function

Xiaoqing Xu [1,2], Xiangsheng Xie[1,3], Hexiang He[4], Huichang Zhuang[1], Jianying Zhou[1,*], Abhilash Thendiyammal[2], Allard P Mosk[2,*]

[1]*School of Physics, Sun Yat-sen University, Guangzhou 510275, China*
[2]*Nanophotonics, Debye Institute for Nanomaterials Science, Utrecht University, P.O. Box 80.000, 3508 TA Utrecht, The Netherlands*
[3]*Department of Physics, College of Science, Shantou University, Shantou, Guangdong 515063, China,*
[4]*School of Physics and Optoelectronic Engineering, Foshan University, Foshan 528000, China*
*\*stszjy@mail.sysu.edu.cn*
*\*a.p.mosk@uu.nl*

We demonstrate a high-speed method to image objects through a thin scattering medium and around a corner. The method employs a reference object of known shape to retrieve the speckle-like point spread function of the scatterer. We extract the point spread function of the scatterer from a dynamic scene that includes a static reference object, and use this to image the dynamic objects. Sharp images are reconstructed from the transmission through a diffuser and from reflection off a rough surface. The sharp and clean reconstructed images from single shot data exemplify the robustness of the method.

## I. Introduction

When light passes through, or reflects off, a turbid medium, it gets scattered and the information it carries is scrambled. This makes looking through diffusing layers or around corners impossible [1]. Scattering is a problem in many imaging scenarios, for instance, the atmospheric agitation in ground-based astronomy, biological tissue in medical imaging and foggy weather in daily life [2-4]. It is in principle possible to retrieve the image of the object obscured by scattering media as the information is not completely lost. Indeed, it has been demonstrated that a speckle pattern propagated through clear space contains enough information to reconstruct the image of an object [5]. Recently, a variety of approaches have been demonstrated to solve this problem, such as time-of-flight imaging [6], time reversal or phase conjugation [7-14], transmission matrix measurement [15-18], wavefront shaping technique [19-23], digital holography [24-27], speckle auto-correlation method [28-34], and other speckle correlation methods [35, 36]. The transmission matrix fully characterizes the effect of a scattering medium. Once the transmission matrix is measured, the image of the hidden object can be reconstructed. Speckle autocorrelation has been shown to be a viable method for non-invasive imaging through uncharacterized opaque media, within the range of the memory effect [37-39]. The method relies on the Gerchberg–Saxton (GS) algorithm to find the phase associated with the power spectrum of the autocorrelation signal, the norm of which is achieved from the autocorrelation of the speckle patterns. In wavefront shaping methods, the incident light wave is modulated with a designed spatial phase to generate a focused wave behind the scattering media. After the wavefront shaping process, objects that are obscured by the scattering media can be imaged. In a previous demonstration, He *et al*. have realized imaging an unknown object through a diffuser by exploiting wavefront shaping with a known object [21]. This method requires an iterative process to shape the speckle pattern to become the reference object at the beginning. Furthermore, the reference object should be removed after the calibration process. More recently, with the knowledge of the point spread function (PSF) measured before data acquisition [40], Zhuang *et al*. showed high speed full-color imaging through a diffuser [41]. Several other deconvolution experiments have been reported [42-44] which require the measurement of PSF in advance. Recently,

Antipa *et al*. demonstrated the reconstruction of light field and digital refocus from the caustics pattern caused by a phase diffuser [45]. These methods require detailed characterization of the scattering medium, using full access to both sides of the medium which may not be available in many practical settings.

In this paper, we present a speckle-imaging method to reconstruct incoherently illuminated objects that are hidden by thin scattering media, requiring only the placement of a reference object of known shape. We use the fact that the transmission pattern due to a single point of the reference object is a high-contrast interference pattern similar to laser speckle [46] as long as the coherence length of the light is longer than the typical difference in path lengths of the scattered light. The transmission pattern of an extended incoherently illuminated object is the intensity sum of many such speckle patterns, each due to a single coherence area of the object. In case the scattering medium is sufficiently thin, due to the optical memory effect [37] the speckle pattern from each coherence area has the same shape and is only displaced. In this case one can regard it as the point spread function (PSF) of the scattering medium, and the observed intensity pattern is the convolution of the PSF and the object's shape. As a result of this convolution the contrast of the observed patterns is relatively low and the statistical properties are different from laser speckle. However, we here use the term speckle in order to describe these patterns conveniently. Our imaging method is based upon retrieving the scatterer's point spread function through a single shot deconvolution algorithm and requires no iteration. Moreover, in a dynamic scene, the reference speckle does not need a separate acquisition process and can be extracted by averaging the speckle patterns. The proposed method is so robust that reconstruction of the image of a hidden object is possible even with a part of the speckle pattern. We extend the proposed method to non-invasively image an object in the reflection from the rough surface of a metal plate. Important future applications of our method may be found in security monitoring and bio-medical imaging such as otoscopy and laryngoscopy.

## II.      Principle

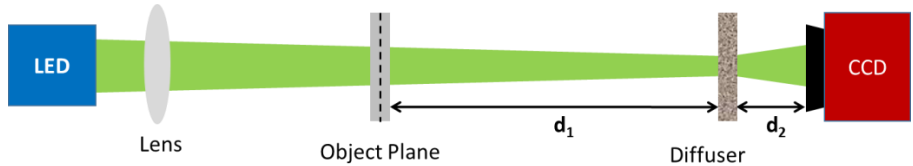

Fig. 1. The schematic of the experimental setup

The schematic of the experimental setup is shown in Fig. 1. A known object (transmittance plate of a letter "H" (Fig. 2(a))) placed in a plane, which we indicate as the object plane, is illuminated by an incoherent beam from a light emitting diode (LED). The light is transmitted through the scattering medium, an optical diffuser, which scrambles the wavefront and a speckle pattern is observed on the CCD (charge coupled device). Within the range of the memory effect, any small tilt in the incident beam results in a tilt of the transmitted speckle [38-40]. Consequently, a point light source in the object plane gives rise to a speckle pattern on the CCD, and by shifting the point source over a small (paraxial) angle one obtains a shifted copy of the same speckle pattern. An incoherently illuminated object hence produces an intensity sum of many copies of the same speckle pattern. Consequently, for objects that have an angular size smaller than the range of the memory effect, the diffuser can be thought of as a lens, though with a speckle pattern as its point spread function (PSF). Under the incoherent illumination, the speckle pattern of the transmittance plate "H" can be expressed as the convolution of the intensity transmittance "H" and this PSF,

$$S_H(x,y) = \iint H(x',y') \times PSF(x-x', y-y')\, dx'dy'$$
$$= H(x,y) * PSF(x,y) \qquad ,(1)$$

Where $S_H$ is the speckle pattern resulting from the known reference object "H" and $*$ denotes the convolution product. When an unknown object "T" is introduced next to the reference object, the speckle pattern $S_{sum}$ on the CCD is the sum of the intensities of the individual speckle patterns:

$$S_{sum} = S_H + S_T = (H+T) * PSF \qquad ,(2)$$

where $S_T$ is the speckle pattern of the unknown object "T". For the convenience, we have removed coordinates $(x, y)$ from the equations.

The relation of two speckle patterns can be exploited to retrieve the image of the unknown object "T". The convolution in the spatial domain is a multiplication in the spatial frequency domain, which yields

$$F\{S_H\} = F\{H\} \times F\{PSF\} \qquad ,(3)$$

$$F\{S_{sum}\} = F\{H+T\} \times F\{PSF\} \qquad ,(4)$$

where "$F$" represents Fourier transform, and $\times$ indicates multiplication. The information of the object "T" can be found by a formal deconvolution as follows,

$$H + T = F^{-1}\left\{\frac{F\{H\} \times F\{S_{sum}\}}{F\{S_H\}}\right\} \qquad ,(5)$$

where "$F^{-1}$" stands for the inverse Fourier transform. In Eq. (5) the division in the frequency domain represents an effective Wiener deconvolution algorithm which is more stable against noise [47]. This procedure amounts to an implicit retrieval of the PSF. In principle the same result is obtained through explicit retrieval of the PSF, which necessitates one extra deconvolution step, and hence is slightly less robust.

### III. Experiments

As shown in the schematic setup (Fig. 1), a commercial LED with a wavelength of 630-650 nm is used as the light source. The light passes through an object (transmittance plate) before impinging on a diffuser (a turbid plastic sheet). A CCD (Basler, dark-60 μm) captures the speckle pattern due to the scattering. The distance from the object plane to the diffuser surface ($d_1$) is 190 mm and that from the CCD to the diffuser surface ($d_2$) is 100 mm. The effective magnification of the scattering lens is $M_{scat} = d_2/d_1$.

The image reconstruction process is depicted in Fig. 2. The speckle pattern of a reference object (letter "H" shown in Fig. 2(a)) is first captured (Fig. 2(b)). A test object, e.g. a letter "T" (Fig. 2(c)), is then added adjacent to the reference object, resulting in a different speckle pattern (Fig. 2(d)). Using the reference object "H" to retrieve the point spread function, the image of the letter "H" and "T" behind the diffuser are recovered according to Eq. (5). The reconstructed result is shown in Fig. 2(e).

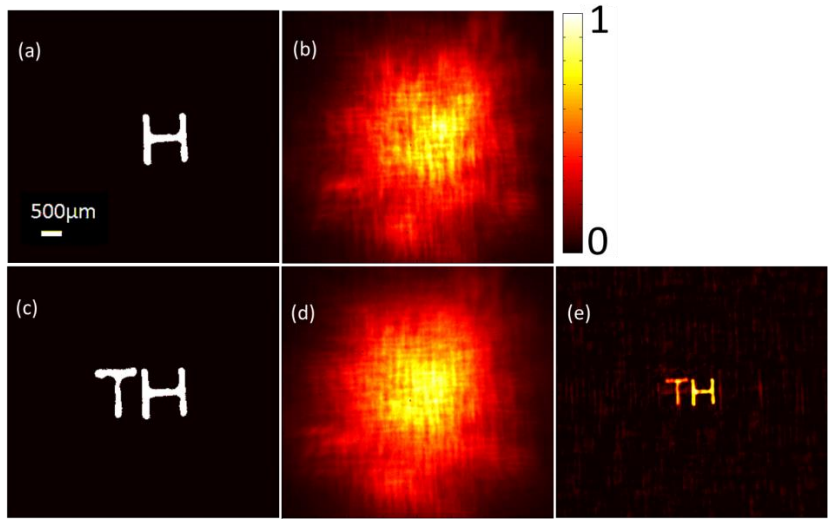

Fig. 2. (a) The reference object "H". The scale bar is 500 µm, which is the same for all images. (b) The speckle pattern of the reference object "H". (c) is the reference object and the unknown object "T" and (d) is the corresponding speckle pattern. (e) is the corresponding reconstructed image.

It is important that the reference object "H" in Eq. (5) is scaled corresponding to the magnification of the system. The reconstruction results are shown versus the scale factor used in the deconvolution process in Fig. 3. If the scaling factor of the reference object is different from the magnification factor of the system (here, $M_{scat}$=0.53), the reconstruction images show a larger amount of background noise and artefacts. When using the correct magnification, the reconstructed images are distinct and clear. In systems where the position $d_2$ of the reference object is unknown, one can generate several reference images with different estimated system magnifications. The clearest reconstruction result indicates the correct system magnification, and thereby yields the position $d_2$ of the reference object. The reconstruction process is performed in parallel on an NVidia GTX 970GPU in approximately 25ms, which is fast enough for real time imaging applications.

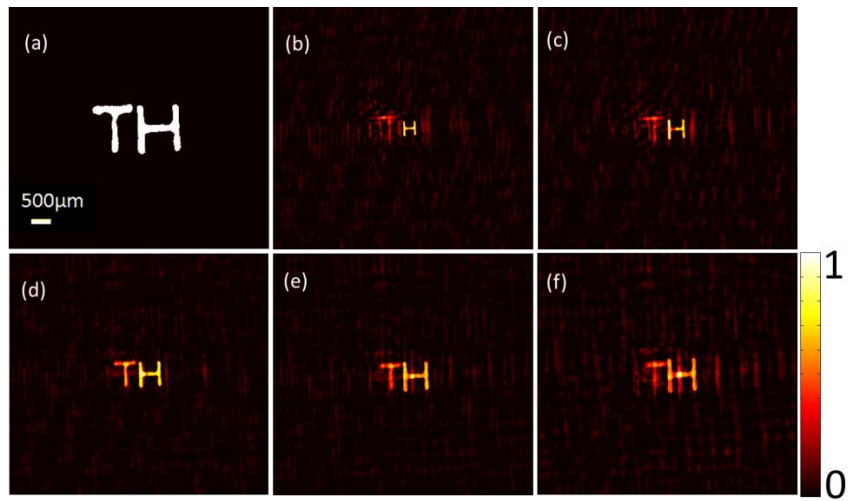

Fig. 3. Reconstructed images of the hidden object "T" with different sizes of the reference object "H": (a) the image of the transmittance plate. The magnification of the "H" used in the deconvolution is 0.3 in (b), 0.4 in (c), 0.53 in (d) (the magnification of the experiment setup), 0.6 in (e), 0.7 in (f), respectively.

The speckle image that we record has holographic properties, in that it does not require the full speckle image to reconstruct an image. Fig. 4 depicts reconstructions based on partial data. The reconstructed results with one-half and one-quarter of the speckle image are shown in Fig. 4(c) and (f), respectively. Even when three quarters of the speckle image is discarded, the only effect on the image is a slight deterioration of the background noise.

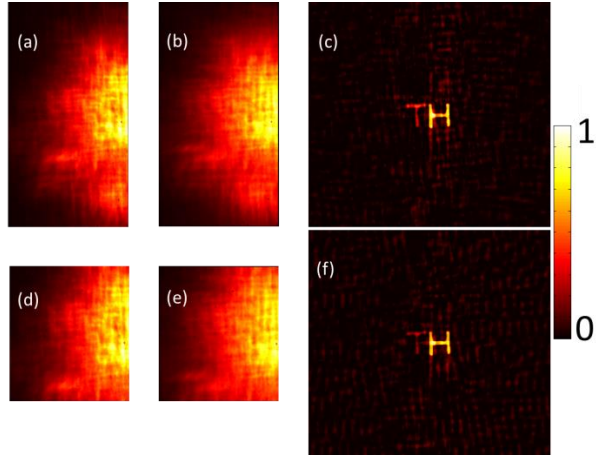

Fig. 4. Reconstructed images from part of the speckle pattern. The half speckle pattern of reference object "H" (a) and the unknown object "T" (b) and the corresponding reconstructed image (c). The quarter speckle pattern of reference object "H" (d) and the unknown object "T" (e) and the corresponding reconstructed image (f).

## IV. Static reference speckle extraction from a dynamic scene and subsequent dynamic image reconstruction

In many speckle-based imaging methods, motion of the test object is a problem. On the contrary, in our deconvolution method it is highly advantageous if the test object is dynamic. In this case the reference pattern is extracted from the time-average speckle pattern of the dynamic scene, and no separate reference measurement is needed.

The experimental setup for dynamic imaging is shown in Fig. 5. A digital projector (with lens removed) is used to generate intensity objects at the plane of its intensity-mode liquid crystal screen. The diffuser (Newport 10° light shaping diffuser) scatters the light from the projector and a lens system composed of a 4x objective lens and a tube lens is used to collect the scattered light for a better signal to noise ratio (SNR) [41]. The collection lens system does not image the diffuser to the CCD. The distance from projector plane to the diffuser surface is 20.7 cm and that from the objective lens to the diffuser surface is 0.7 cm. The total magnification of the imaging system was measured to be M= 0.1 using the method of Fig. 3.

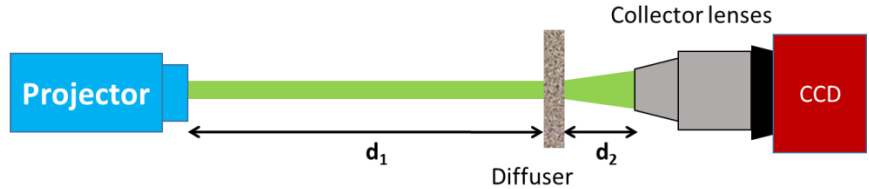

Fig. 5. The experiment setup for imaging a dynamic object through a diffuser

The projector displays an image of a small disk moving around a bifurcation (Fig. 6(a)), which mimics the situation of a cell moving around a static object. The CCD captures a series of speckle images of the scene with different positions of the disk. By averaging the series of speckle patterns, the speckle pattern of the reference object is obtained (Fig.6. (b)). With the speckle pattern of the reference object, the image of the moving object is retrieved from the corresponding speckle pattern by applying Eq. (5). Fig. 6.(f), (g), and (h) show the reconstructed images of the moving object at different locations. A video of dynamic imaging through the diffuser can be found in the supplementary file. The number of frames that is needed to obtain the speckle pattern of the reference object depends on the complexity of the dynamic scene. For the model in Fig. 6, 360 frames were used, however an acceptable SNR was already obtained with 100 images.

The key of successful implementation of the dynamic scene based reference speckle extraction method is the existence of a static reference object on the scene. The reconstruction algorithm only needs to know the shape of the reference object. A possible example is the use of fluorescence imaging near an implanted object [48]. The implanted object remains static while fluorescently marked cells are flowing. If we know the shape of the implant, it is possible to exploit this imaging method.

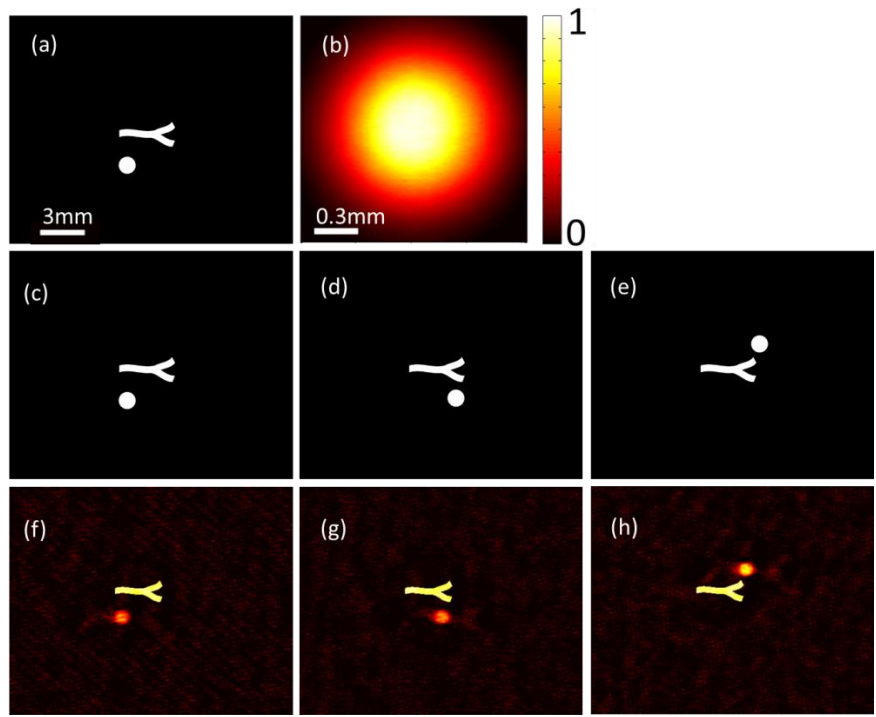

Fig. 6. Imaging moving object (a circle displayed on the projector) through the diffuser. (a) The reference object (b) the approximated speckle pattern of the reference object, which is mean value of the 360 speckle images of the moving object. (c), (d), (e) are the original images of the reference object and the moving object and (f), (g), (h) are the corresponding reconstructed images. The scale bar in (a) is 3mm and it is the same in the (c), (d) and (e). The scale bar in (b) is 0.3mm and it is the same in the (f), (g) and (h).

## V.  Looking around corners using nonspecular reflections

In many situations such as laryngoscopy or otoscopy it is useful to extract an image from a reflection in a nonspecular surface. Here we demonstrate that PSF retrieval recovers images from the reflection of a nonspecular test sample, a rough metal plate. We use a 4f conjugated setup which images the surface of a mirror to that of the metal plate as shown in Fig. 8. First, an unknown object "T" (Fig. 9(a)) is illuminated with a LED light and the transmitted light is incident on the metal plate (Fig. 9 (c)) at an angle. The speckle pattern due to the scattered light is acquired by the CCD and is shown in Fig. 9(b). Afterwards, another beam is used to illuminate the reference object "H" at normal incidence, such that its diffraction pattern on the metal plate overlaps with that of the unknown object "T". The corresponding speckle pattern is captured on the CCD as shown in Fig. 9(e).

Compared to the previous cases (section 3 and 4), here the known and unknown objects are presented separately in space. Therefore, the Eq. (5) can be modified as

$$T = F^{-1}\left\{\frac{F\{H\} \times F\{S_T\}}{F\{S_H\}}\right\} \quad .(6)$$

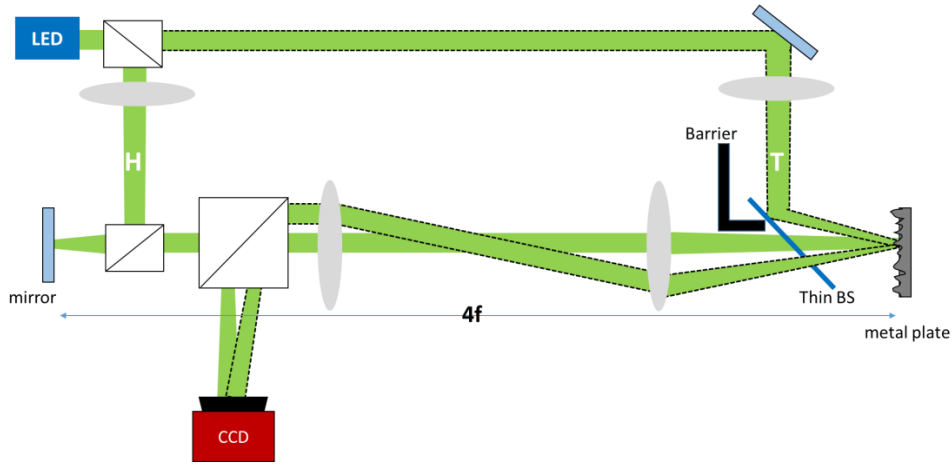

Fig. 8. Imaging an object around a metal plate. The mirror and the metal plate are conjugated by the 4f systems, of which the focus length of the lens is 150 mm. The dotted line represents the light which illuminates the unknown object and reflected by the metal plate.

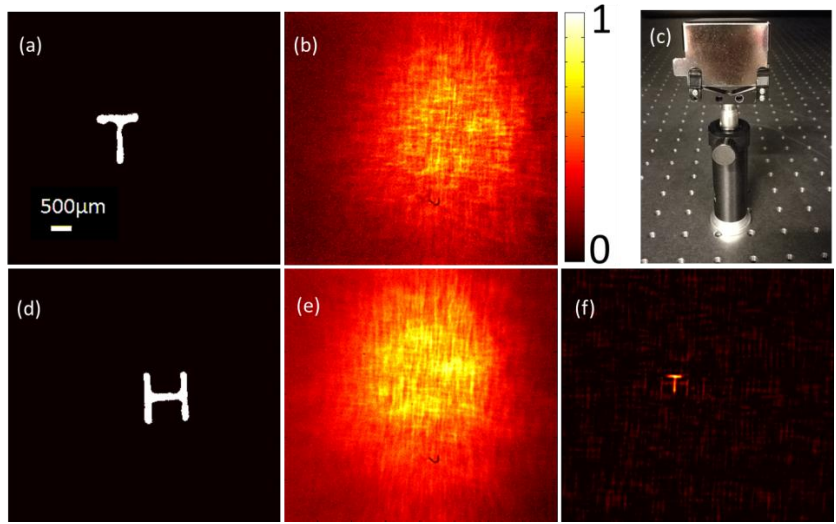

Fig. 9. Imaging in the reflection form a rough surface (a) The unknown object "T". The scale bar is 500μm, which is the same for the else images. (b) The speckle pattern of the object "T". (c) is a metal plate of which the surface is rather rough. (d) is the reference object "H" and (e) is the corresponding speckle pattern. (f) is the reconstructed image of the unknown object "T".

The images of two speckle patterns and the prior information of the reference object are used to retrieve the unknown object, even though the unknown object is present at an angle (approx. 1 degree) that is different from that of the reference. Fig. 9(f) shows the retrieved image with a magnification factor of 0.345. It is to be noted that, here the reference object is no longer required to be close to the unknown object, which really improve the practicability. It is found that the method is limited to a small incident angle as the reflected light should also go through the lens system and reach the CCD. If the angle is too large, no light will enter the lens system. This limitation of our apparatus could be improved by tilting the lens system to align with the near-specular reflected light.

## VI. Discussion

Our method uses the speckle-like pattern produced by a known reference object to retrieve the point spread function of the scattering medium. The reference object does not need to be point-like and it does not need to be present separately or to be removed before an image of another object is reconstructed. Compared to other methods like wavefront shaping and transmission matrix, the speckle point spread function retrieval method is robust, high speed and requires a much simpler setup as it can work with incoherent light. Compared to time-of-flight methods, our method is much easier and can form an image in a single exposure [6]. Since the method retrieves the true point spread function, it is in principle possible to use digital propagation to image objects that are not in the plane of the reference object, making this a 3D capable imaging method.

However, point spread functions only work within the angular range of optical memory effect (known as the isoplanatic patch in lens design) and consequently the corresponding field of view (FOV) is rather small. The FOV may be enlarged by shrinking and scanning the illumination light that was performed in a previous work [21]. The same can be achieved as follows: Firstly, the image of an unknown object can be reconstructed with a prior known object and then the object of this restored image can be treated as another known reference object and other objects around it can be recovered by tilting the illumination to them. Likewise, once images of other objects are reconstructed, their corresponding objects will become the known reference objects and more images of other nearby objects can be restored thus the FOV can be improved.

## VII. Conclusion

A method is presented to image objects behind or around a scattering medium based on retrieval of the point spread function by speckle deconvolution. The proposed method relies on the prior knowledge of the shape of a reference object, and makes use of the memory effect to faithfully reconstruct the image of objects near the reference object. The recovery approach is single-shot and high speed, with calculations performed on a millisecond time scale. Moreover, the speckle from the static reference object can be extracted from a dynamic scene. Thus the proposed method has the potential to image moving objects behind a scattering medium. This principle could give rise to applications in tissue optics, when reference objects whose shape is known, for example through non-optical imaging, are present. Moreover, we can use the same principle to image through reflections in non-specular surfaces, where the PSF is measured at a different angle from the object, opening up new possibilities for imaging in confined spaces.

## Acknowledgements

A. P. Mosk acknowledges a Vici grant from the Netherlands Organization for Scientific Research (NWO). This work is also supported by the Chinese National Natural Science Foundation (11534017 & 61575223). We also acknowledge the State Key Laboratory of Optoelectronic Materials and Technologies (Sun Yat-sen Unversity). X. Xu acknowledges support by China Scholarship Council. We thank Jeroen Bosch, Siddharth Ghosh, Pritam Pai for helpful discussion.